# Responsive Equilibrium
# for Self-Adaptive Ubiquitous Interaction

Massimiliano Dal Mas

*me @ maxdalmas.com*

**ABSTRACT**
This work attempts to unify two domains: the Game Theory for cooperative control systems and the Responsive Web Design, under the umbrella of crowdsourcing for information gain on Ubiquous Sytems related to different devices (as PC, Tablet, Mobile, …).
This paper proposes a framework for adapting DOM objects components for a disaggregated system, which dynamically composes web pages for different kind of devices including ubiquitous/pervasive computing systems. It introduces the notions of responsive web design for non-cooperative Nash equilibrium proposing an algorithm (RE-SAUI) for the dynamic interface based on the game theory.

**ACM Classification Keywords**
H.5.m. Information interfaces and presentation (e.g., HCI): Miscellaneous.

**INTRODUCTION**
On the Web efficient dynamic resource provisioning algorithms are necessary to the development and automation of responsive webdesign in a dynamic environment. The main goal of these algorithms is to offer services that satisfy the requirements of individual users – that can dynamically define the webpage – while guaranteeing at the same time an efficient utilization of the devices (browsers) used – according to their technical requirements. The goal of this work is to overcome the current limitations of responsive web design (e.g. CSS3 and HTML5 media queries) when considering dynamic web pages constructed by a users that need to be displayed in different browsers for different devices with different orientation. To implement the proposed model we propose a dynamic algorithm based on a mathematical formulation of the allocation problem that maximizes the DOM objects used. The solution of this model allows us to obtain a performance achievable by any browsers allocation algorithm.

———

Following we introduce the "responsive webdesign" and the DOM "forest" structure of a webpage. The DOM objects of the Nash equilibrium tree according to their location in the forest is used to define the proposed algorithm "Responsive Equilibrium for Self-Adaptive Ubiquitous Interaction" (RE-SAUI).

**RESPONSIVE NON-COOPERATIVE NASH EQUILIBRIUM**
Recently, the emergent concept of the "responsive webdesign" approaches to website design keeps up with the endless new resolutions and devices.
The goal of "responsive webdesign" is to build web pages that detect the visitor's screen size and orientation to change the layout accordingly. Instead of having to build a special version of a website for every kind of device - which often requires to rewrite HTML code - developers can repurpose the same HTML code using a mix of fluid layouts, CSS media queries and scripts that can reformat web pages automatically.

To perform that a whole new way of thinking about design was proposed. But responsive Web design cannot only consider adjustable screen resolutions and automatically resizable images when faced website is dynamically build according to the user wishes.
Dynamic websites have to "respond" to the wishes of users and to the features of different devices. Those two class requirements can be in contrast in some cases performing two different strategies. Thinking at "class requirements" as agents: a combination of strategies is such that each agent's component strategy is that agent's best reply to the other agent's best reply to it. A combination of strategies for different agents is defined on a "non-cooperative Nash equilibrium combination" if and only if no single agent $A_i$ could unilaterally increase his utility by choosing a different strategy from among the strategies available to him, while keeping the strategies of the other agents $A_{n-i}$ fixed [1].

**DOM "FOREST" STRUCTURE**
According to the W3C "The Document Object Model is a platform and language neutral interface that will allow programs and scripts to dynamically access and update the content, structure and style of documents." (http://www.w3.org/DOM)

The HTML DOM views an HTML document as a tree-structure (node-tree) defined by nodes that can be accessed through the tree. The node tree contents can be modified or deleted, and new elements can be created.

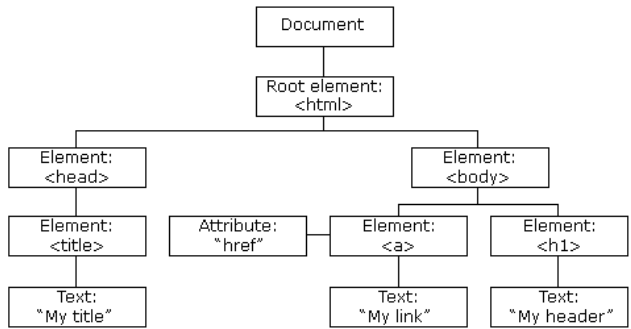

**Figure 1** The tree starts at the root node and branches out to the text nodes at the lowest level of the tree. The node tree shows the set of nodes, and the connections between them.

```
<html>
<head>
<title>My title</title>
</head>
<body>
<h1>My header</h1>
<a href="uri">My link</a>
</body>
</html>
```
**Code 1** The corresponding HTML code for the tree displayed in Figure 1.

When a browser renders an HTML page, the HTML is downloaded into local memory and parsed using the DOM to construct the internal data structures employed to display the page in the browser window. The DOM is also used by JavaScript to examine HTML page respect the state of the browser. DOM APIs are used by JavaScript code to inspect or modify a webpage. According to the W3C a webpage has a logical structure which is very much like a "forest", which can contain more than one DOM tree. The HTML root element of the "forest" serves as the root of the DOM tree for the webpage displayed (see Figure 2). From the root element it is possible do define a tree path that connect the DOM objects that dynamically construct the webpage according to the user wishes and the browser constraints (according to the user and browser requirements). Consequently the single tree defined in the forest represents the "strategy" followed by a browser to display a webpage composed by the DOM object defined in the tree path.

**DOM TREES AS NASH EQUILIBRIUM**
In many scenarios the page construction design is not defined a priori, but turn out from the interactions of distinct independent selfish DOM objects selected by the user or dynamically selected by an application. Nowadays DOM objects should be selected even by each browser according to his requirement (as display definition and its orientation) to be visualized on a constructing page. As example let us consider a user that composing his page on a PC wants to have the same component (DOM objects) on a Mobile (see Figure 3).

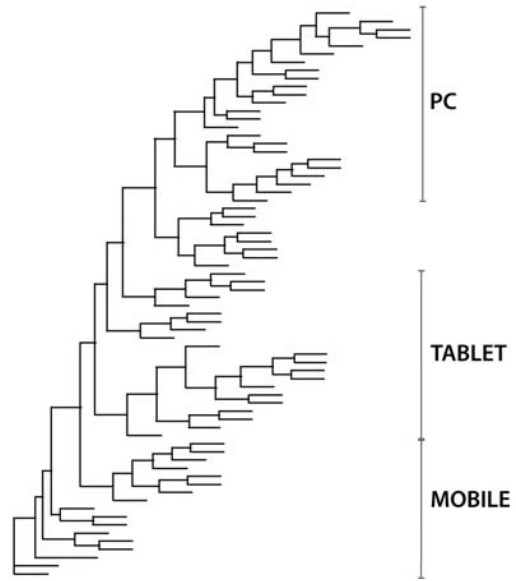

**Figure 2** DOM "forest" simplified representation containing more DOM trees according to the device (e.g. Pc, Tablet, Mobile)

We consider a DOM tree design of the construction of a webpage build by browsers with independent selfish DOM objects.
The DOM objects tree ($T$) is composed dynamically according to the choice of the user and the browser $i$ requirements: so consequently the root-leaf of the DOM object tree represents the browser. Each construction of a dynamic page (the browser goal) is displayed by a browser according to the minimum page cost. We can consider the game theory [4] as natural frame for the interaction of independent selfish DOM objects (chosen by the users) and the browsers. A Nash Equilibrium (NE) is a set of users' choices, such that none of them has an incentive to deviate unilaterally. For this reason the corresponding trees of DOM objects can be considered a stable configuration.

Anyway, Nash equilibrium in a tree equilibrium – and generally in game – could give a more expensive solution respect the optimal one, thought as a unified optimal solution where the webpage is predefined a priori (not automatically constructed by a user). The design costly tree is due to mainly by the lack of cooperation among tree DOM objects, that reflect a selfish use of DOM object.

We are now ready to describe the DOM trees representing Nash equilibrium. The graphs were constructed as instances of geodetic graphs by Blokhuis and Brouwer [2]. It was showed in [3] that every Nash equilibrium forms a tree.
We define a direct graph tree $T = (V;E)$, represented by DOM objects, with Vertex ($V$) and Edges ($E$). In our discussion, when we refer to a point or a line, we often mean the corresponding vertex or player. Every player is mean the corresponding vertex or player. Every player is represented by a tree DOM objects where the node are the DOM objects that "play" to be shown on the constructed web page by the browser (that follow a "strategy"). Taking for instance the example depicted in Figure 1: we have $V=11$ and $E=V-1=10$. We classify the vertices, as DOM objects, of the equilibrium tree according to their location in the forest. We refer to the vertices at maximum depth as vertices in the Boundary level [3].

**OVERLAY TREE OF DOM OBJECTS**

In this work we overcome the limitation of considering a non cooperative environment by suggesting a new overlay tree game, that we call "Responsive Equilibrium for Self-Adaptive Ubiquitous Interaction" (RE-SAUI) game. In our proposal DOM objects are defined by an algorithm that merge individual and cooperative interests in a unique tree DOM structure. The function of the cost for each DOM object is a merge of the path cost toward it (the selfish component) and the tree cost to build a web page, which represents the cooperative component.

**RESPONSIVE EQUILIBRIUM**

This section illustrates the proposed "Responsive Equilibrium for Self-Adaptive Ubiquitous Interaction" (RE-SAUI), motivating the reason to introduce such model.

The RE-SAUI algorithm occurs in a directed tree $T = (V,E)$, where each edge $e$ has a nonnegative cost $c_e$, and each browser $i \in I = \{1, 2, \ldots k\}$ is identified with a root-leaf pair $(r_i, l_i)$ that define the page-construction path. Every browser $i$ picks a "strategy" as page-construction path $P_i$ from its root to its leaf and create a tree $(V, \cup_i P_i)$ with a total cost $C$ defined by (1) which will be referred to as *page* cost.

$$(1) \quad C = \sum_{e \in \cup_i P_i} c_e$$

We consider the page-construction path $P_i$ as the "strategy" chosen by a browser $i$ to construct a webpage according to its requirement (e.g. for a tablet we can consider the display resolution and orientation). The total page cost $C$ is shared among the DOM objects depicted in the "forest" (see *DOM "FOREST" STRUCTURE*) in the following way: let $x_e$ denote the number of overlay page-construction tree paths that go through overlay edge $e$ connecting the DOM objects; if edge $e$ lies in $x_e$ of the chosen tree path for the page construction then the proportional cost of such browser "strategy" is defined by (2).

$$(2) \quad \pi_e = c_e / x_e$$

The total cost function of the DOM objects $Z_i$ that browser $i$ aims to minimize is given by (3).

$$(3) \quad Z_i = \sum_{e \in P_i} \pi_e + \delta \sum_{e \in \cup_j P_j} c_e$$

The first term in (3) considers the independent *selfish* essence of every DOM object representing the cost for browser $i$ to acquire a DOM object belonging to the chosen tree path for the page-construction. The second term in (3) represents the total tree cost for the page-construction (i.e., the *page* cost), where the parameter $\delta$ is used to weight the constitutive components. A different explanation of total cost function (3) is likewise achievable: DOM objects can be considered as fully selfish (depicted in the first term) – as they can be freely chosen by the user to compose a webpage – while the browser imposes an awareness on their use for the tree page construction (depicted in the second term).

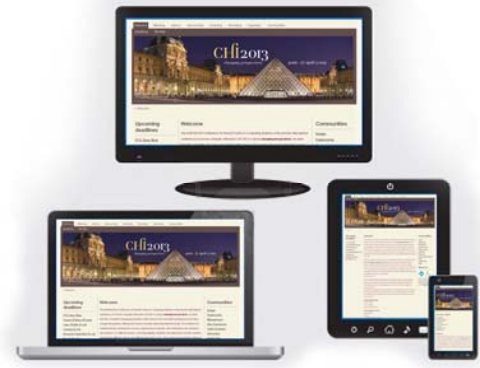

**Figure 3:** Responsive "equilibrium" webdesign (RE-SAUI) for a webpage displayed according to its resolution and orientation (self-adaptive ubiquitous interaction).

The proposed approach uses an optimization on the page construction choices (see Figure 3) for every change in the page-construction tree (either in the topology, links cost, browser location etc…) without having to consider a big Integer Linear Programming [1] problem but letting the browsers to solve the problem in a distributed way that converge to a good and stable solution. Note that for $\delta = 0$ total cost function (3) corresponds to that defined by the Shapley tree design game proposed in [6], that can be thought as a specific condition of the game proposed here. Moreover, in our game, browsers need only a limited amount of information, which is exactly equal to that of the Shapley tree design game [6].

The RE-SAUI game is related to the "potential games" class of the game theory that was determined in [7]. The "potential games" are described by a vector $\Phi(P)$ for the tree path page-construction that measures the cost difference saved by any browser if it is the only browser to take a different path from the other browsers. We define the vector of the browsers' strategies as $P = (P_1, P_2, \ldots P_k)$. The strategy performed by all other browsers than $I$ is defined as $P_{-i}$, and $P'i \neq P_i$ is an alternative strategy for some browser $i$.

---

[1] **Linear Programming** (LP) is a specific case of mathematical programming (mathematical optimization) process of finding the extreme values (maximum and minimum values) of a function for a region defined by linear relationships. In a linear program, the objective function and the constraints are linear relationships, the effect of changing a variable is proportional to its magnitude. When the variables take integer values the problem is called an Integer Linear Programming (ILP). ILP's occur frequently because options must be chosen from a finite set of alternatives where decisions are discrete (as yes/no). While LP can be solved efficiently in the worst case, ILP are in many cases NP-hard (when considering bounded variables).

Such that a website that is been viewed by $k$ browsers can be defined by a potential function, $\Phi(P)$, where for any browser $i$ we have $\Phi(P) - \Phi(P') = Z_i(P) - Z_i(P')$

On "potential games" exist at least a strategy $P$ that minimizes $\Phi(P)$ considered as a pure Nash equilibrium [6] where the best response dynamics converge. So the potential function (3) characterizes the RE-SAUI game.

$$(3) \quad \Phi(P) = \sum_{e \in E} \sum_{x=1}^{x_e} \frac{c_e}{x} + \delta \sum_{e \in \cup_j P_i} c_e$$

The cost incurred by a browser $i$, $Z_i$, considers the cost of the total tree page-construction (multiplied by the parameter $\delta$). Different levels of cooperative awareness or DOM object cooperation is taken into account changing the value of $\delta$.

The RE-SAUI design problems are NP-hard, however, the proposed Mixed Integer Linear Programming (MILP) formulations can be solved to the optimum for realistic size instances in reasonable time.

Introducing a term related to the tree page-construction cost in the total cost function (3) for each browser, we expect that browsers should design better trees path minimizing (3).

## IMPLEMENTATION RE-SAUI

We define here a possible implementation on the best dynamic interface response that converge to a Nash equilibrium. The simple algorithm allows each browser to improve its cost function on building a page. Minimizing objective function (3) and assuming that other DOM objects are not changing their strategies we can obtain the best response strategy for a DOM object

We consider a graph tree $T = (V,E)$, where each DOM object $e$ never has a negative cost $c_e$. Let us consider a vector of browsers' strategies $P = (P_1, P_2, \ldots P_k)$, where $P_i \subset O$ is a set of DOM objects that connect the root of the DOM tree with the leaf displayed in the browser page $(p_i, t_i)$. The set of edges used by all browsers except $i$ is defined by $P_{-i}$. So all links that are not used by the set of all browsers except $i$ is defined by the set $E - (P_{-i})$. The number of times that edge $e$ lies in the paths chosen by such browser (i.e. all DOM objects except DOM object $i$) is written as $k_e$

The best response to all other strategies $P_{-i}$ can be determined by Algorithm 1. According to that algorithm the cost incurred by browser $i$ to choose the DOM objects on the tree is set equal to the link weights, according to the total cost function of the DOM objects $Z_i$ defined by (3).

Then, the best response is given by the minimization of the user's cost that correspond to the shortest path on the DOM tree. The browser chooses a random tree path if multiple shortest paths have the same cost.

1: **if** $e \in P^{-i}$ **then**
2:     Set $Cost(e) = c_e / (k_e + 1)$
3: **else**
4:     Set $Cost(e) = c_e \cdot (\delta + 1)$
5: **end if**
6: Compute link costs $Cost(e)$ for the the "root-leaf cost path"
7: Set $P_i$ equal to the set of links included in $Cost(e)$

**Algorithm 1 Pseudo-code for the Responsive Equilibrium for Self-Adaptive Ubiquitous Interaction (RE-SAUI).**

## CONCLUSION

In this paper we investigated the design problem for responsive on interactive web pages in terms of deciding the use of the DOM objects to be visualized in a browser page, the assignment of browsers to access DOM objects, the optimal combination of DOM objects in order to obtain a self-adaptive ubiquitous interaction on browsers towards a responsive equilibrium. To this end, we proposed a "Responsive Equilibrium for Self-Adaptive Ubiquitous Interaction" (RE-SAUI) design models, based on mathematical programming, which take into account the individual requirements of DOM objects chosen by a user, the connectivity between them (depicted as nodes) and the browser. Our proposed models can solve two different problems: the minimization of the DOM tree initialization cost while ensuring full coverage of all DOM objects relative to the browsers used, and the maximization of the DOM objects used by choosing the DOM objects respect the browsers' specification.

**Massimiliano Dal Mas** is an engineer working on webservices and is interested in knowledge engineering. His interests include: user interfaces and visualization for information retrieval, automated Web interface evaluation and text analysis, empirical computational linguistics, text data mining, knowledge engineering and artificial intelligence. He received BA, MS degrees in Computer Science Engineering from the Politecnico di Milano, Italy. He won the thirteenth edition 2008 of the CEI Award for the "best degree thesis" with a dissertation on "Semantic technologies for industrial purposes" (Supervisor Prof. M. Colombetti). In 2012, he received the "best paper award" at the IEEE Computer Society Conference on Evolving and Adaptive Intelligent System (EAIS 2012) at Carlos III University of Madrid, Madrid, Spain. In 2013, he received the "best paper award" at the ACM Conference on Web Intelligence, Mining and Semantics (WIMS 2013) at Universidad Autónoma de Madrid, Madrid, Spain. His paper at 2013 W3C Workshop on Publishing using CSS3 & HTML5 has been appointed as "position paper", Paris, France.